# Revisiting longitudinal optical modes in materials with plasmon and plasmon-like absorptions – SrTiO$_3$ and β-Ga$_2$O$_3$

Thomas G. Mayerhöfer[a,b,*] and Jürgen Popp[a,b]

[a] *Leibniz Institute of Photonic Technology (IPHT), Albert-Einstein-Str. 9, D-07745 Jena, Germany*
[b] *Institute of Physical Chemistry and Abbe Center of Photonics, Friedrich Schiller University, Jena, D-07743, Helmholtzweg 4, Germany*

[*]*Corresponding author. Tel.: +49 (0)3641/948348; Fax: +49 (0)3641/206399; E-mail address: Thomas.Mayerhoefer@ipht-jena.de (T.G. Mayerhöfer)*

**ABSTRACT**

We investigate exemplary the longitudinal optical (LO) mode order in compounds with a plasmon or plasmon-like phonon mode and additional phonon modes. When the oscillator strength of the plasmon or plasmon-like mode is gradually increased, a reordering of the modes takes place. Since it is not possible in crystals with orthorhombic or higher symmetry that a LO mode crosses a transverse optical (TO) mode's position, this reordering takes place via mode hybridization. During this mode hybridization, the plasmon or plasmon-like LO mode gradually becomes the originally higher situated LO mode while the latter morphs into the former. As a consequence, an inner (LO-TO) and an outer (TO-LO) mode pair is formed. This process continues until the LO oscillator strength is so high that all other phonons are inverted and form LO-TO pairs within the outer TO-LO mode pair of the plasmon or plasmon-like mode. These insights can be readily transferred to other semiconductors or many mode materials with reststrahlen bands and allow simple mode assignments. These mode assignments will help to understand the nature of surface modes of structured layers of these materials for application of surface plasmon polariton and surface phonon polaritons based metamaterials.

# 1. Introduction

Plasmonic metals like Au and Ag have their plasma frequencies in the visible spectral range and are therefore plagued by strong optical losses in the infrared spectral region. Alternatives are on the one hand transparent conductive oxides like $In_{2-x}Sn_xO_3$,[1] $ZnO$[2-3] or $Ga_2O_3$[4-5] or other doped semiconductors like $InAs$[6] and $InAsSb$[7] which have their plasma frequencies located in the infrared and support in structured form also surface plasmon polaritons (SPPs). As an alternative, it is also possible to use surface phonon polaritons (SPhPs) to generate surface enhanced infrared absorption instead. Corresponding materials gain increasingly attention over the last years due to their low optical losses despite of high enhancement factors comparable to those achievable with the classic plasmonic metals.[8-11] To generate SPhPs in a wider spectral range, the use of materials with broad reststrahlen regions is of advantage, therefore SiC,[8, 11-12] cBN[13] and hBN[14] have been considered up to now among a few other materials. Recently, also $SrTiO_3$ has gained attraction, since its reststrahlenband is exceptionally broad and promises wide tunability of the SPhPs correspondingly.[15] A reststrahlenband is generally characterized by high reflectance close to unity between its transversal optical (TO) mode frequency and its longitudinal optical (LO) mode position, between which the real part of the dielectric function is usually negative. This situation is comparable to that of a metal below its plasma frequency. Correspondingly, about a century ago, it was common to talk about "metallic stripes" when referring to strong absorptions.[16-18] For $SrTiO_3$, the situation is somewhat more complex, since it features 3 different absorptions which have their TO mode positions within the reststrahlenband.[15, 19-21] The TO-LO pairs have been assigned simply consecutively, i.e. with increasing wavenumber, the first TO mode is assumed to be followed by its LO mode and so forth. Indeed, there is a deep cleft in the reststrahlenband of $SrTiO_3$ corresponding to a region where the real part of the dielectric function is positive, and it seems to be plausible, at least for the third TO mode, which has a wavenumber higher than the cleft, to be followed by its LO mode. For the second mode, which is constituted only by a small dip in the reflectance, the assignment is not so obvious. It is correct to state that from a mathematical point of view each TO mode must be followed by a LO mode (at least for crystals with orthorhombic symmetry or higher),[22] but as shown by Gervais it is possible that a weak mode is inverted and has an LO mode at lower wavenumber than its TO counterpart.[23] If there is a strong TO mode at lower wavenumber, and the weak mode is located within the range in which the real part of the dielectric function is negative, then the weak mode is inverted and the LO mode belonging to the strong TO mode has a wavenumber higher than the weak TO mode. Correspondingly, both modes establish an inner (weak mode) and an outer TO-LO pair (strong mode).

Recently, we have established a connection between the oscillator strengths $S^2 = Nq^2/\mu\varepsilon_0$ of the TO and the LO modes, where $N$ is the number of oscillator per unit volume, $q$ the (effective) charge, $\mu$ the reduced mass and $\varepsilon_0$ the permittivity of vacuum. For materials with only one phonon mode (or with one phonon mode per principal dielectric function in optically uniaxial and orthorhombic materials), the corresponding relation is given by,[24]

$$S^2 = S_{TO}^2 = \varepsilon_\infty^2 S_{LO}^2, \qquad (1)$$

when the classical damped harmonic oscillator model is assumed. Correspondingly, for the dispersion relations,

$$\begin{aligned}\varepsilon_r(\tilde{\nu}) &= \varepsilon_\infty + \frac{S^2}{\tilde{\nu}_{TO}^2 - \tilde{\nu}^2 - i\gamma\tilde{\nu}} \quad &\text{(I)} \\ \varepsilon_r^{-1}(\tilde{\nu}) &= \varepsilon_\infty^{-1} - \frac{S^2/\varepsilon_\infty^2}{\tilde{\nu}_{LO}^2 - \tilde{\nu}^2 - i\gamma\tilde{\nu}} \quad &\text{(II)}\end{aligned}, \qquad (2)$$

holds. Here, $\varepsilon_\infty$ is a static background resulting from modes situated at such high wavenumber that they are separated by transparency region from the modes under discussion, with an approximately constant and real value of the dielectric function represented by $\varepsilon_\infty$. $\tilde{\nu}_{TO}$ and $\tilde{\nu}_{LO}$ are the mode positions and $\gamma$ is the damping constant. For systems with more than one oscillator, eqn. (1) does no longer hold for the LO modes due to mode coupling. Nevertheless, in particular for weak modes, the oscillator strengths are still comparably to those for uncoupled modes and allow in context with the mode positions an unambiguous assignment of the modes. Providing such an assignment is one of the goals of this work. Since the reststrahlenband in SrTiO$_3$ is plasmon-like, we will transfer the insights gained from the analysis of its modes to analyze a transparent conducting oxide having also more than one phonon mode. Here, we focus on the modes with transition moments parallel to the *b*-axis of monoclinic Ga$_2$O$_3$, which has just recently been analyzed,[5, 25] to detail our extended approach. The results will hopefully be helpful to extend the insights gained and the approach employed to any non-metallic material of interest for tuning SPhPs and SPPs, so that the modes generated in corresponding structures can be understood and conforming design rules can be worked out.

**2. Theory, results and discussion**

Eqn. (2), (I) can be extended to more than one oscillator:

$$\varepsilon_r = \varepsilon_\infty + \sum_{j=1}^{N} \frac{S_j^2}{\tilde{v}_{TO,j}^2 - \tilde{v}^2 - i\gamma_j \tilde{v}}. \tag{3}$$

Eqn. (3) has been used to investigate a reflectance spectrum of SrTiO$_3$, and the oscillator parameters have been determined.[19] While we are aware of the fact that eqn. (3) is not able to capture the subtle features caused by the coupling between the very strong oscillator at about 88 cm$^{-1}$ with the other two oscillators, the simple model above is fully sufficient to guide us with regard to the assignment of the LO modes. Eqn. (3) will be flanked by its counterpart for the LO-modes,

$$\varepsilon_r^{-1} = \varepsilon_\infty^{-1} - \sum_{j=1}^{N} \frac{S_j^2/\varepsilon_\infty^2}{\tilde{v}_{LO,j}^2 - \tilde{v}^2 - i\gamma_j \tilde{v}}, \tag{4}$$

where we use the Lyddane-Sachs-Teller (LST) relation, to compute $\tilde{v}_{LO,j}^2$ from the relation:[24]

$$\tilde{v}_{LO,j}^2 = \tilde{v}_{TO,j}^2 + S_j^2/\varepsilon_\infty. \tag{5}$$

The positive sign in eqn. (5) must be replaced by its negative counterpart if a much stronger oscillator $j$-1 is present so that $\tilde{v}_{TO,j-1} < \tilde{v}_{TO,j} < \tilde{v}_{LO,j-1}$.[23-24, 26]

As mentioned above, eqn. (4) is exact only for one oscillator, because it does not account for the strong coupling present for the LO-modes. Nevertheless, a comparison between the negative imaginary part of inverted dielectric function $-\text{Im}\{1/\varepsilon_r\}$ from eqn. (3) and $-\text{Im}\{\varepsilon_r^{-1}\}$ based on eqn. (4) helps to make the mode assignments and reveals information about the strength and the nature of the coupling. For SrTiO$_3$, we will first make the assumption that only the first oscillator is present. Accordingly, we set $S_2$ and $S_3$ initially zero and then increase their values at the same time gradually until both reach their final values. We calculate $-\text{Im}\{1/\varepsilon_r\}$, determine the location of the three maxima $\tilde{v}_{LO,j}$ and the peak values multiplied by $\tilde{v}_{LO,j}$: $-\tilde{v}_{LO,j}\,\mathrm{Im}\{1/\varepsilon_r\}$. The connection between the latter value and the actual LO oscillator strength, if there is only one oscillator, is given by:

$$S_{LO} = \sqrt{-\tilde{v}_{LO}\gamma\,\mathrm{Im}\{1/\varepsilon_r\}}\,. \tag{6}$$

If eqn. (6) is applied for cases with more than one oscillator to determine $S_{LO,j}$ and compared with those obtained by using eqn. (7),

$$\varepsilon_r^{-1} = \varepsilon_\infty^{-1} - \sum_{j=1}^N \frac{S_{LO,j}^2}{\tilde{v}_{LO,j}^2 - \tilde{v}^2 - i\gamma_j \tilde{v}}\,, \tag{7}$$

to actually fit the negative inverse of the dielectric function $-\tilde{v}_{LO,j}\,\mathrm{Im}\{1/\varepsilon_r\}$, the differences can be considerable as consequence of LO mode coupling.

In the following, we determine $-\tilde{v}_{LO,j}\,\mathrm{Im}\{1/\varepsilon_r\}$ in dependence of the plasma frequency instead. We then compare $-\tilde{v}_{LPP,j}\,\mathrm{Im}\{1/\varepsilon_r\}$ and $-\tilde{v}_{LPP,j}\,\mathrm{Im}\{\varepsilon_r^{-1}\}$, keeping in mind that for the second term $\tilde{v}_{LO,j}$ is different and given by eqn. (5). The corresponding results are depicted in Figure 1.

If the oscillator strengths of mode 2 and 3 are both set to zero, the LO mode position would be located at 684 cm$^{-1}$. A gradual increase of the squared oscillator strengths of mode 2 and 3 leads to a shift of this LO mode to higher wavenumbers which would be unexpected in the uncoupled case where this mode is supposed to not change its position. It could therefore be inferred, that this LO mode must be assigned to mode 3, the wavenumber position of which is supposed to increase with the wavenumber. However, the increase would be too strong and comparison of the intensities clearly shows that the LO mode with the highest wavenumber is mostly constituted by mode 1, so it must be assigned to this mode. At the same time, the wavenumber position of LO mode 3 starts at its TO wavenumber and decreases with increasing oscillator strength, which is a clear sign that $\tilde{v}_{TO,1} < \tilde{v}_{TO,3} < \tilde{v}_{LO,3}$. The decrease of the wavenumber position can be described to a good approximation by assuming uncoupled LO-modes. In addition, its intensity is not much different from that of the uncoupled case. Therefore, its assignment is unambiguous. The remaining LO mode shows some interesting properties. Despite a comparably high TO oscillator strength, its LO mode strength is weak and more than two orders of magnitude lower at the expense of the mode strength of mode 1. Correspondingly, its LO mode position is even at full TO oscillator strength only a few cm$^{-1}$ lower than its TO mode, because it would otherwise approach the TO position of the first mode, which is generally not possible. The small decrease of the LO mode position of mode 2 is a consequence of the high oscillator strength of mode 1.

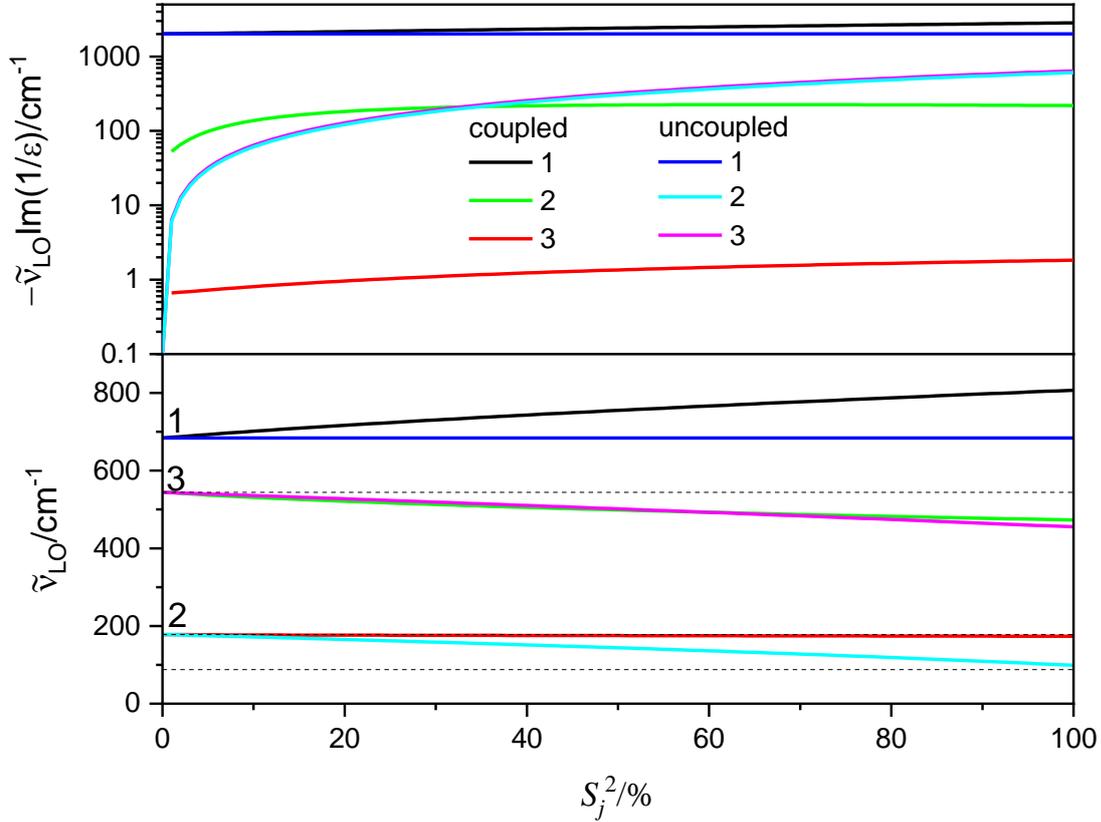

*Figure 1: Upper panel: negative imaginary part of the inverse of the model dielectric function ($1/\varepsilon_r$) and the model inverse dielectric function ($\varepsilon_r^{-1}$) times the LO wavenumber position as a function of the percentual oscillator strength $S_j^2$. Lower panel: LO wavenumber positions, i.e. the wavenumber locations of the maxima of the negative imaginary part of the inverse of the model dielectric function ($1/\varepsilon_r$) and the model inverse dielectric function ($\varepsilon_r^{-1}$). The dashed lines indicate the TO oscillator positions.*

Overall, also this assignment is unambiguous, not only because it is the last remaining mode, but also because its LO mode position starts at the TO mode position. The fact that it transfers oscillator strength to mode 1 and, correspondingly, stays close to the TO position becomes important in the following example, since SrTiO$_3$'s first mode is with its low TO mode position and high TO oscillator strength very similar to a plasmonic mode, even if the underlying absorption is not at all plasmonic in nature. For the following discussion, however, the absorption mechanism is not of importance. Following ref. [25] "the LPP mode coupling for A$_u$ symmetry (modes in Ga$_2$O$_3$) is trivial and equivalent to any other semiconductor material whose unit eigendisplacement vectors are all parallel and/or orthogonal". In principle we agree with the authors, but think we can amend their analysis of the $Ga_2O_3$ $A_u$ modes and that provided recently in ref. [5] with some valuable information. These might be helpful also for the analysis of similar materials.

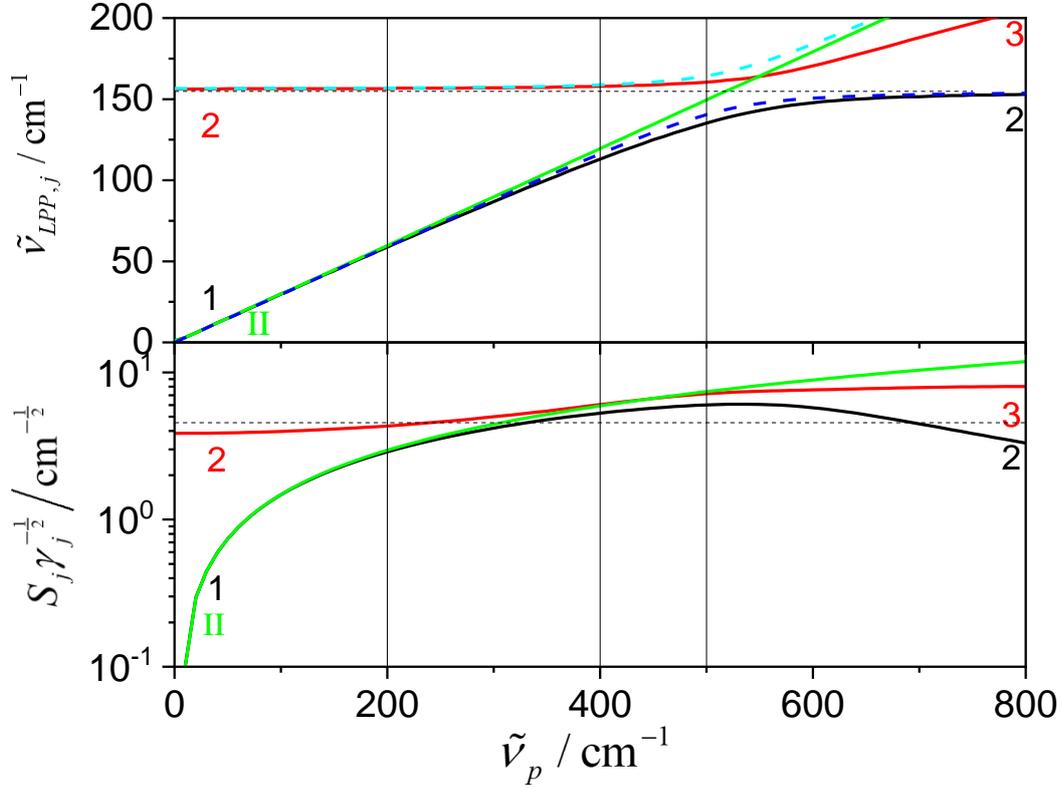

*Figure 2: Upper panel: LO wavenumber positions, i.e. the wavenumber location of the maxima of the negative imaginary part of the inverse of the model dielectric function ($1/\varepsilon_r$) and the model inverse dielectric function ($\varepsilon_r^{-1}$). The thin black dashed line indicates the TO oscillator position of the first phonon mode (mode 2). The blue and the turquoise dashed lines indicate the LPP mode positions $\tilde{N}^2_{LO\pm}$ calculated from eqn. (10). Lower panel: square root of the negative imaginary part of the inverse of the model dielectric function ($1/\varepsilon_r$) and the model inverse dielectric function ($\varepsilon_r^{-1}$) times the LP wavenumber position as a function of the plasma frequency (cf. eqn. (6)). The thin black dashed line represents the intensity of mode 2 according to $\sqrt{-\tilde{\nu}_{LPP,2} \operatorname{Im}\{1/\varepsilon_r\}}$.*

For the LPP mode analysis, we keep all parameters of the 4 $A_u$ phonon modes fixed at the values provided in [5] and change the oscillator strength of the plasmon, i.e. the plasma frequency $S_p(=\tilde{\nu}_p)$, gradually, like this was done in [5], so that the results can be readily compared. Transferred to the previous example of $SrTiO_3$, this would mean that we freeze the oscillator parameter of mode 2 and 3 at their determined values and increase the oscillator strength of mode 1 step by step. In contrast to this first mode of $SrTiO_3$, a plasmon mode has its TO position at zero wavenumber ("free electrons"). To increase the plasmon frequency beyond the TO wavenumber of the highest phonon mode for β-$Ga_2O_3$, it would be necessary to choose $\tilde{\nu}_p^2 > \tilde{\nu}_{TO,4}^2 \varepsilon_\infty$.[27] Computing $\tilde{\nu}_p$ using the values provided by ref. [5], this means $\tilde{v}_p > 1279$ cm$^{-1}$, but before increasing the plasma frequency to this value we first inspect closer in Figure 2 how the LPP mode positions change with smaller increases of the plasma frequency.

In the lower part in Figure 2 the changes of the quantity $\sqrt{-\tilde{v}_{LPP,j}\,\text{Im}\{1/\varepsilon_r\}}$ are depicted (black and red lines). In order to better understand their changes with increasing plasma frequency, we additionally computed $\sqrt{-\tilde{v}_{LPP,1}\,\text{Im}\{\varepsilon_r^{-1}\}}$ for the uncoupled plasmon according to eqn. (2) (II), assuming that $\varepsilon_\infty$ is equal to $\varepsilon_0 = 11.251$ (cf. [5]). The idea behind has been developed in [24] and assumes that as long as the LO- or LPP-mode situated at lower wavenumbers is well-separated from the other modes, it is a non-coupling mode and the influence of the other modes can be captured by the (approximately) constant and real value of the dielectric function between this mode and the next one. For a very weak plasmon mode this value would simply be the (real) value of the dielectric function at zero wavenumber. Indeed, it can be seen in the lower part of Figure 2 that the curve of the uncoupled plasmon (green curve) agrees well with that of the coupled plasmon up to about $\tilde{v}_p \leq 200$ cm$^{-1}$. The same is true for the LPP-mode position shown in the upper part of Figure 2, where the uncoupled plasmon LO-mode position was calculated by $\tilde{v}_{LO,1} = \sqrt{\tilde{v}_p^2/\varepsilon_0}$ (cf. eqn. (5)). Again, beyond $\tilde{v}_p = 200$ cm$^{-1}$, deviations began to emerge between the black and the green curve. Going back to the lower part of Figure 2, it is obvious from inspecting $\sqrt{-\tilde{v}_{LPP,2}\,\text{Im}\{1/\varepsilon_r\}}$ for mode 2 and comparing it with $\sqrt{-\tilde{v}_{LPP,2}\,\text{Im}\{\varepsilon_r^{-1}\}}$ (thin black dashed line), that mode 2 is also not strongly coupled, neither to mode 1 nor to the other modes. Above $\tilde{v}_p = 200$ cm$^{-1}$, however, $\sqrt{-\tilde{v}_{LPP,2}\,\text{Im}\{1/\varepsilon_r\}}$ increases up to about $\tilde{v}_p \leq 400$ cm$^{-1}$ where it agrees for nearly 100 cm$^{-1}$ with $\tilde{v}_{LO,1} = \sqrt{\tilde{v}_p^2/\varepsilon_0}$. The agreement is not perfect, but good enough to conclude that in this region, mode 2 has actually changed into the plasmon mode and the former plasmon mode has become mode 2. Since then $\tilde{v}_p^2 > \tilde{v}_{TO,2}^2 \varepsilon_{0,2}$, where $\varepsilon_{0,2}$ would be the real value of the dielectric function, would there be no first phonon mode, the real part of the dielectric function is now negative, therefore the LO-mode position for this mode is now *smaller* than its TO-mode position, just like this is the case for mode 2 and 3 in SrTiO$_3$. With further increase of $\tilde{v}_p$, the LO-mode position will approach the TO-mode position as can be seen in the upper part of Figure 2, a situation which is very comparable to that of mode 2 in SrTiO$_3$, where the LO-mode position is only a few cm$^{-1}$ below the TO-mode position, despite of being a comparably strong mode. The variation of the LO-mode position can be modelled assuming mode coupling. A corresponding formula has been introduced by Gervais.[23] Based on the condition $\varepsilon' = 0$, by negligence of damping the following relation results for the situation at hand:

$$\tilde{N}^2_{LPP\pm} = \frac{1}{2}\left\{\tilde{\nu}^2_{LPP,1} + \tilde{\nu}^2_{LPP,2} \pm \left[\left(\tilde{\nu}^2_{LPP,1} - \tilde{\nu}^2_{LPP,2}\right)^2 + 4\frac{S_1^2 S_2^2}{\varepsilon^2_{0,1}}\right]^{\frac{1}{2}}\right\}, \qquad (8)$$

Here, the $\tilde{\nu}_{LPP,j}$ represent the uncoupled LO-mode resonance positions, whereas $\tilde{N}_{LPP\pm}$ are the LO-mode resonance positions for the coupled oscillators. Note that $\varepsilon^2_\infty$ has to be replaced by $\varepsilon^2_{0,1}$ and that the $S_j$ are the TO oscillator strengths.

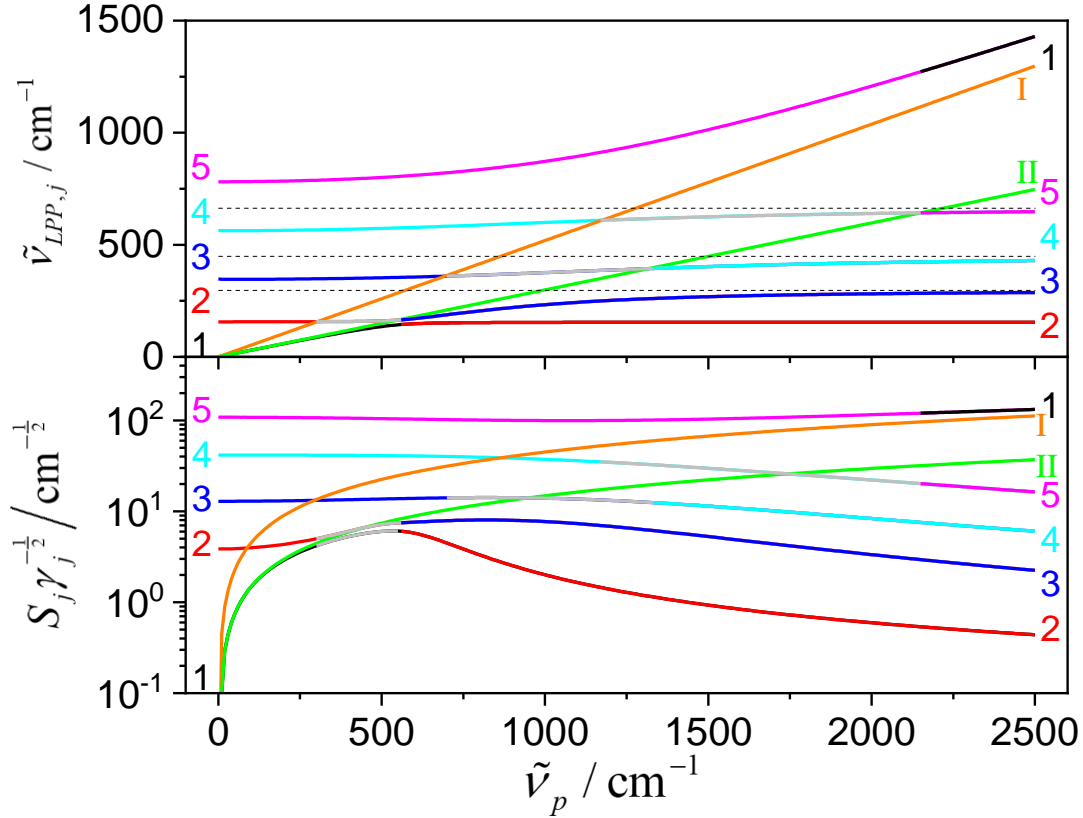

*Figure 3: Upper panel: LO wavenumber positions, i.e. the wavenumber locations of the maxima of the negative imaginary part of the inverse of the model dielectric function (1/ε$_r$). The thin black dashed lines indicate the TO oscillator position of the four phonon modes. The orange and the green lines specify the plasmon mode position of an uncoupled mode. Lower panel: square root of the negative imaginary part of the inverse of the model dielectric function (1/ε$_r$) and the model inverse dielectric function (ε$_r$ $^{-1}$) multiplied by the LP wavenumber position as a function of the plasma frequency (cf. eqn. (6)). The green and the orange line represent the same quantity for uncoupled plasmons screened by $\varepsilon_{0,1}$ (green line) and $\varepsilon_\infty$ (orange line) following eqn. (2) (II).*

From the upper part of Figure 2 it is obvious that eqn. (8) can well describe the actual LPP mode positions (turquoise and blue dashed lines) up to the point where $\tilde{\nu}_p \leq 400$ cm$^{-1}$. Beyond this point, the plasmon mode begins to stronger hybridize with the second phonon mode and the first phonon mode is largely described by the black curve. This stronger hybridization is also reflected in the crossing between the green and the red line, which should not take place according to eqn. (8), since the turquoise line stays always above the green line. This crossing is therefore an indication that afterwards the red line represents phonon mode 2 (LPP mode 3). Accordingly, in Figure 3, the colors of the modes have been switched and a grey zone has been introduced. This grey zone begins at the curve defined by $\tilde{\nu}_{LO,1} = \sqrt{\tilde{\nu}_p^2 / \varepsilon_\infty}$, which is just approximately an indicator for the start of the hybridization of the modes. Obviously, as soon as the mode positions begin to increase with increasing plasma frequency in the left upper part of Figure 3, hybridization sets in. It is certainly correct to say that for crystal symmetries up to orthorhombic no LPP mode crosses the line of a TO mode. Instead, through hybridization, such crossings are avoided, but nevertheless the LPP modes finally end up having a lower wavenumber than the corresponding TO modes. Therefore, if compared with the situation in SrTiO$_3$, at high mode strengths of the plasmon mode, the corresponding LPP mode is the one with the highest wavenumber. This is also clearly seen in Figure 3 where for large plasma frequencies the highest mode increases both in strength as well as with regard to the mode position like the plasmon mode in the uncoupled model (orange line).

## 3. Summary and Conclusion

In summary, we have discussed the LO mode order in SrTiO$_3$ and the LPP mode order of the modes of A$_u$ symmetry of β-Ga$_2$O$_3$. If the latter is highly doped, then both set of modes have in common, that the mode with the lowest wavenumber is so strong that the real part of the dielectric function is in wide ranges negative and its highest LO or LPP mode is situated at wavenumbers beyond all other modes. Those other modes have LO or LPP positions that are lower than the corresponding TO modes. Based on the discussion we provided it should be easy to determine the mode order in other materials with strong low lying TO modes that are plasmon-like or plasmons.

**Table of Contents entry**

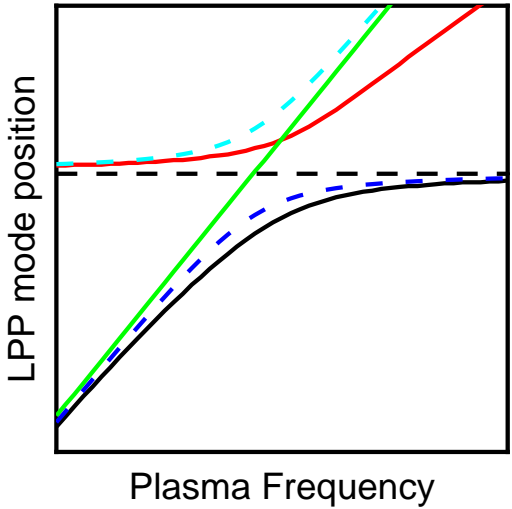

**Keywords:** Keywords: Dispersion analysis; Dielectric loss function; Inverse dielectric function; Dielectric function modelling; TO-LO rule